\documentclass[pra,showpacs,showkeys,amsfonts,amsmath,twocolumn]{revtex4}
\usepackage{bm}
\usepackage{graphicx}
\RequirePackage{mathptm}

\numberwithin{equation}{section}

\newcommand{\sa}[2]{\sigma_{#1}^{#2}}

\newcommand{\WW}[6]{\begin{cases}
#1 ,&#2\\
#3 ,&#4\\
#5 ,&#6
\end{cases}}
\newcommand{\ro}{\varrho}
\newcommand{\ras}{\varrho_{\mr{as}}}

\newcommand{\al}{\alpha}

\newcommand{\g}{\gamma}
\newcommand{\ga}[1]{\gamma_{#1}}
\newcommand{\Ga}[1]{\Gamma_{#1}}
\newcommand{\Om}[1]{\Omega_{#1}}

\newcommand{\re}{\mathrm{Re}\,}
\newcommand{\I}{\openone}
\newcommand{\conj}[1]{\overline{#1}}
\newcommand{\ket}[1]{|{#1}\rangle}
\newcommand{\bra}[1]{\langle {#1} |}

\newcommand{\ptr}[1]{\mathrm{tr}_{#1}}

\newcommand{\mr}[1]{\mathrm{#1}}

\begin{document}
\title{Delayed birth of distillable entanglement in the evolution of bound entangled states}
\author{{\L}ukasz Derkacz}
\affiliation{Opera Software International AS\\ Oddzia{\l} w Polsce
\\
ul. Szewska 8, 50-122 Wroc{\l}aw, Poland}
\author{Lech Jak{\'o}bczyk
\footnote{E-mail addres: ljak@ift.uni.wroc.pl} }
 \affiliation{Institute of Theoretical Physics\\ University of
Wroc{\l}aw\\
Plac Maxa Borna 9, 50-204 Wroc{\l}aw, Poland}
\begin{abstract}
The dynamical creation of entanglement between three-level atoms
coupled to the common vacuum is investigated. For the class of bound
entangled initial states we show that the dynamics of closely
separated atoms generates stationary distillable entanglement of
asymptotic states. We also find that the effect of delayed sudden
birth of distillable entanglement occurs  in the case of atoms
separated by a distance comparable with the radiation wavelength.
\end{abstract}
\pacs{03.67.Mn; 03.65.Yz; 42.50.-p}
\keywords{three-level atoms; entanglement production; distillable entanglement}
 \maketitle

\section{Introduction}
Dynamical creation of entanglement by the indirect interaction between
otherwise decoupled systems has been studied by many researchers
mainly in the case of two-level atoms interacting with the common
vacuum. When the two atoms are separated by a distance small
compared to the radiation wave length $\lambda$, there is a substantial
probability that a photon emitted by one atom will be absorbed by
the other and the resulting process of photon exchange produces
correlations between the atoms. Such correlations may cause that
initially separable states become entangled (see e.g. \cite{J,
FT1,FT2,JJ1}). The case of three-level atoms is very interesting for many reasons.
First of all, in a system of coupled multi - level atoms having closely lying
energy states and interacting with the vacuum, quantum interference
between different radiative transitions can occur, resulting in
coherences in a system which are known as \textit{vacuum - induced
coherences}. Beside the usual effects such as collective damping and dipole-dipole
interaction involving non-orthogonal transition dipole moments \cite{Ag, FS},
radiative coupling can produce
here a new interference effect in the spontaneous emission.
This effect manifests by the cross
coupling between radiative transitions with \textit{orthogonal}
dipole moments \cite{AP} and is strongly dependent on the relative
orientation of the atoms \cite{EK,SE}. All such collective
properties of the system influence the quantum dynamics, which can
significantly differ from a corresponding single atom dynamics.
On the other hand, the theory of
entanglement between the pairs of such atoms is much more complex then in the case of qubits.
As is well known, there is no simple necessary and sufficient condition of entanglement, since
the Peres-Horodecki separability criterion \cite{P, HHH1} only shows that
the states which are not positive after partial transposition (NPPT states) are entangled.
But there can exist entangled states which are positive after this operation and such states
are not distillable \cite{HHH2}. Hence all entangled states can be divided into two classes, one
contains free entangled states that can be distilled using local operations and classical
communication (LOCC) and the other consist of bound entangled states for which no LOCC strategy can
be used to extract pure state entanglement. Since many effects in quantum information exploit pure
maximally entangled states, only distillable states can be directly used for quantum communication.
Contrary, nondistillable states  states involve some kind of irreversibility: we need pure
entanglement to create them, but no pure entanglement can be obtained back from them \cite{YHHS}.
\par
The dynamics of entanglement between three-level atoms with
vacuum-induced coherences was previously investigated mainly in the
context of creation or degradation of NPPT states \cite{DJ1,DJ2}. In
particular it was shown that for small distance between the atoms
the system decays to a stationary state which can be entangled, even
if the initial state was separable \cite{DJ1}. On the other hand, if
the distance is comparable to the radiation wavelength, the dynamics
brings all initial states into the asymptotic state in which both
atoms are in their ground states but still there can be some
transient entanglement between the atoms \cite{DJ2}.
\par
In the present paper, we investigate the dynamics of bound entangled
initial states. There are some results concerning decoherence and
disentanglement of bound entangled states \cite{Sun}, but here we
focus on the process of dynamical creation of distillable
entanglement due to the collective damping and cross coupling
between the three-level atoms. For the specific bound entangled
initial state and small interatomic distance we show that the
asymptotic state is both entangled and distillable. Thus we obtain a
stationary free entanglement. The same result is also valid for other initial
states including separable states. So this dynamics of three - level atoms distinguishes
 distillable states, since all  nontrivial asymptotic entangled states are also distillable.
For larger distances, the dynamics of
the bound entangled initial state is very peculiar: the system very quickly
disentangle and only after some finite time there  suddenly appears
a distillable entanglement. (The similar phenomenon of delayed
sudden birth of entanglement was observed in the case of two-level
atoms \cite{FTd}). So also in this situation the physical process of
spontaneous emission can create some transient distillable
entanglement out of the initially prepared bound entanglement.
\section{Mixed - state entanglement and distillation}
\subsection{Distillability of entanglement}
Distillability of mixed entangled state $\ro$ is the property that
enables to convert $n$ copies of $\ro$  into less number of $k$
copies of maximally entangled pure state  by means of LOCC \cite{B}.
It is known that all pure entangled states can be reversibly
distilled \cite{BB} and any mixed two-qubit entangled state is also
distillable \cite{HHH3}. In general case, the following necessary
and sufficient condition for entanglement distillation was shown in
Ref. \cite{HHH2}: the state $\ro$ is distillable if and only if
there exists $n$ such that $\ro$ is $n$ - copy distillable i.e.
$\ro^{\otimes n}$ can be filtered to a two-qubit entangled state.
This condition is however hard to apply, since conclusions based on
a few copies may be misleading \cite{W}. More practical but not
necessary condition is based on the reduction criterion of
separability \cite{C}. The criterion can be stated as follows: if a
bipartite state $\ro$ of a compound system $AB$ is separable, then
\begin{equation}
\ro_{A}\otimes \I-\ro\geq 0\quad\text{and}\quad \I\otimes
\ro_{B}-\ro\geq 0\label{red}
\end{equation}
where
$$
\ro_{A}=\ptr{B}\,\ro,\quad \ro_{B}=\ptr{A}\,\ro
$$
As was shown in Ref.\cite{HH},  any state that violates (\ref{red})
is distillable, so if
\begin{equation}
\ro_{A}\otimes \I-\ro\not\geq 0\quad\text{or}\quad\I\otimes
\ro_{B}-\ro\not\geq 0\label{reddist}
\end{equation}
 the state $\ro$ can be distilled. The condition (\ref{reddist})
is easy to check and we will use it in our discussion of dynamical
aspects of distillability.
\subsection{Peres-Horodecki criterion and bound entanglement}
To detect entangled states of two qutrits, we apply Peres-Horodecki
criterion of separability \cite{P, HHH1}. From this criterion
follows that any state $\ro$ for which its partial transposition
$\ro^{\mr{PT}}$ is non - positive (NPPT state), is entangled. One
defines also  \textit{ negativity} of the state $\ro$ as
\begin{equation}
N(\ro)=\frac{||\ro^{\mr{PT}}||_{\mr{tr}}-1}{2}\label{neg}
\end{equation}
$N(\ro)$ is equal to the absolute value of the sum of the negative
eigenvalues of $\ro^{\mr{PT}}$ and is an entanglement monotone
\cite{VW}, however it cannot detect entangled states which are
positive under partial transposition (PPT states). Such states exist
\cite{H} and as was shown in Ref. \cite{HHH2},  are not
distillable. They are called \textit{bound entangled PPT states}. Up
to now, it is not known if there exist bound entangled NPPT states
\cite{HHHH}.
\par
To detect some of bound entangled PPT states we can use the
realignment criterion of separability \cite{Ch,Rud}. The criterion
states that for any separable state $\ro$ of a compound system, the
matrix $R(\ro)$ with elements
\begin{equation}
\bra{m}\otimes\bra{\mu}R(\ro)\ket{n}\otimes\ket{\nu}=
\bra{m}\otimes\bra{n}\,\ro\,\ket{\mu}\otimes\ket{\nu}\label{realig}
\end{equation}
has a trace norm not greater then $1$. So if the \textit{
realignment negativity}\, defined by
\begin{equation}
N_{R}(\ro)=\max\,\left(0,\frac{||R(\ro)||_{\mr{tr}}-1}{2}\,\right)\label{rneg}
\end{equation}
is greater then zero, the state $\ro$ is entangled. In the case of
two qubits, the measure (\ref{rneg}) cannot detect all NPPT states
\cite{Rud1}, but for larger dimension the criterion is capable of
detecting some bound entangled PPT states \cite{Ch}.

\section{Time evolution of three-level atoms}
To study the dynamics of entanglement between three-level atoms
we consider the model introduced by Agarwal and Patnaik \cite{AP}. We start
with the short description of the model.
 Consider two identical three - level atoms (
$A$ and $B$) in the $V$ configuration. The atoms have two near -
degenerate excited states $\ket{1_{\al}},\; \ket{2_{\al}}$ ($\al
=A,B$) and ground states $\ket{3_{\al}}$. Assume that the atoms
interact with the common vacuum and that transition dipole moments
of atom $A$ are parallel to the transition dipole moments of atom
$B$. Due to this interaction, the process of spontaneous emission
from two excited levels to the ground state  take place in each
individual atom but a direct transition between excited levels is
not possible. Moreover, the coupling between two atoms can be
produced by the exchange of the photons, but in such atomic system
there is also possible the
radiative process in which atom $A$ in the excited state
$\ket{1_{A}}$ loses its excitation which in turn excites atom $B$ to
the state $\ket{2_{B}}$. This effect manifests by the cross coupling
between radiation transitions with orthogonal dipole moments.  The
evolution of this atomic system can be described by the following
master equation \cite{AP}
\begin{equation}
\frac{d\ro}{dt}=i\,[H,\ro]\, + \,(L^{A}+L^{B}+L^{AB})\ro\label{me}
\end{equation}
where
\begin{equation}
\begin{split}
H&=\sum\limits_{k=1}^{2}\Om{k3}\,(\sa{k3}{A}\sa{3k}{B}+\sa{k3}{B}\sa{3k}{A})\\[1mm]
&+\sum\limits_{\al=A,B}\Om{vs}
(\sa{23}{\al}\sa{31}{\neg\al}+\sa{32}{\al}\sa{13}{\neg\al})
\label{hamiltonian}
\end{split}
\end{equation}
and for $\al=A, B$
\begin{equation}
L^{\al}\ro=\sum\limits_{k=1}^{2}\ga{k3}\,\left(\,2\sa{3k}{\al}\ro\sa{k3}{\al}-\sa{a3}{\al}
\sa{3k}{\al}\ro-\ro\sa{k3}{\al}\sa{3k}{\al}\right)  \label{genA}
\end{equation}
moreover
\begin{equation}
\begin{split}
L^{AB}\ro &=\hspace*{3mm}\sum\limits_{k=1}^{2}\sum\limits_{\al=A,B}\Ga{k3}\,(\,
2\sa{3k}{\al}\ro\sa{k3}{\neg\al}-\sa{k3}{\neg\al}\sa{3k}{\al}\ro-\ro\sa{k3}{\neg\al}\sa{3k}{\al})\\[1mm]
&\hspace*{3mm}+\sum\limits_{\al=A,B}\Ga{vc}(\,2\sa{31}{\al}\ro\sa{23}{\neg\al}
-\sa{23}{\neg\al}\sa{31}{\al}\ro-
\ro\sa{23}{\neg\al}\sa{31}{\al}\\[1mm]
&\hspace*{20mm}+2\sa{32}{\al}\ro\sa{13}{\neg\al}-\sa{13}{\neg\al}\sa{32}{\al}\ro
-\ro\sa{13}{\neg\al}\sa{32}{\al}\,)
\label{genAB}
\end{split}
\end{equation}
In the equations (\ref{hamiltonian}), (\ref{genA}) and (\ref{genAB}), $\neg\al$ is $A$
for $\al=B$ and $B$ for $\al=A$, $\sa{jk}{\al}$ is the transition
operator from $\ket{k_{\al}}$ to $\ket{j_{\al}}$  and the
coefficient $\ga{j3}$ represents the single atom spontaneous - decay
rate from the state $\ket{j}$ ( $j=1,2$ ) to the state $\ket{3}$.
Since the states $\ket{1_{\al}}$ and $\ket{2_{\al}}$ are closely
lying, the transition frequencies $\omega_{13}$ and $\omega_{23}$
satisfy
$$
\omega_{13}\approx\omega_{23}=\omega_{0}
$$
Similarly, the spontaneous - decay rates
$$
\ga{13}\approx\ga{23}=\g
$$
The coefficients $\Ga{j3}$ and $\Om{j3}$ are related to the coupling
between two atoms and are the collective damping  and the dipole -
dipole interaction potential, respectively. The coherence terms
$\Ga{vc}$ and $\Om{vc}$ are cross coupling coefficients, which
couple a pair of orthogonal dipoles. Detailed analysis shows the cross coupling between two
atoms strongly depend on the relative orientation of the atoms and there are
such configurations of the atomic system that $\Ga{vc}=\Om{vc}=0$ and the other configurations
for which $\Ga{vc}\neq 0,\, \Om{vc}\neq 0$.
Moreover, all the coupling
coefficients are small for large distance $R$ between the atoms and tend to zero for
$R\to\infty$. On the other hand, when $R\to 0$, $\Om{13},\, \Om{23}$ and $\Om{vc}$
diverge, whereas
$$
\Ga{13},\, \Ga{23}\to \gamma,\quad\text{and}\quad \Ga{vc}\to 0
$$
\par
The time evolution of the initial state of the system is given
by the semi-group $\{ T_{t}\}_{t\geq 0}$ of completely positive mappings acting on density
matrices, generated by the hamiltonian (\ref{hamiltonian}) and the dissipative part $L^{A}+L^{B}+L^{AB}$.
The properties of the semi-group crucially depends
on the distance $R$ between the atoms and the geometry of the system. Irrespective to the geometry,
when $R$ is large compared to the radiation wavelength, the semi-group $\{ T_{t}\}_{t\geq 0}$
is uniquely relaxing with the asymptotic state $\ket{3_{A}}\otimes\ket{3_{B}}$. Thus, for any initial state
its entanglement asymptotically approaches $0$. But still there can be some transient entanglement between the atoms.
On the other hand,
in the strong correlation regime (when $R\to 0$), the semi-group is not uniquely relaxing and the
asymptotic stationary states are non-trivial and depend on initial conditions. The explicit form
of the asymptotic state $\ras$ for any initial state $\ro$ with matrix elements $\ro_{kl}$
(with respect to the canonical basis) was
found in Ref.\cite{DJ1}. It is given by
\begin{equation}
\ras=\begin{pmatrix} 0&0&\hspace*{2mm}0&0&0&\hspace*{2mm}0&\hspace*{2mm}0&\hspace*{2mm}0&\hspace*{2mm}0\\
0&0&\hspace*{2mm}0&0&0&\hspace*{2mm}0&\hspace*{2mm}0&\hspace*{2mm}0&\hspace*{2mm}0\\
0&0&\hspace*{2mm}x&0&0&\hspace*{2mm}z&-x&-z&\hspace*{2mm}w\\
0&0&\hspace*{2mm}0&0&0&\hspace*{2mm}0&\hspace*{2mm}0&\hspace*{2mm}0&\hspace*{2mm}0\\
0&0&\hspace*{2mm}0&0&0&\hspace*{2mm}0&\hspace*{2mm}0&\hspace*{2mm}0&\hspace*{2mm}0\\
0&0&\hspace*{2mm}\conj{z}&0&0&\hspace*{2mm}y&-\conj{z}&-y&\hspace*{2mm}v\\
0&0&-x&0&0&-z&\hspace*{2mm}x&\hspace*{2mm}z&-w\\
0&0&-\conj{z}&0&0&-y&\hspace*{2mm}\conj{z}&\hspace*{2mm}y&-v\\
0&0&\hspace*{2mm}\conj{w}&0&0&\hspace{2mm}\conj{v}&-\conj{w}&-\conj{v}&\hspace*{2mm}t
\end{pmatrix}\label{roas}
\end{equation}
where
\begin{equation}
\begin{split}
&x=\frac{1}{8}\,(\ro_{22}+2\ro_{33}+\ro_{44}+2\ro_{77}-2\,\re \ro_{24}-4\,\re \ro_{37})\\
&z=\frac{1}{4}\,(\ro_{36}-\ro_{38}-\ro_{76}+\ro_{78})\\
&w=\frac{1}{4}\,(\ro_{26}+\ro_{28}+2\ro_{39}-\ro_{46}-\ro_{48}-2\ro_{79})\\
&y=\frac{1}{8}\,(\ro_{22}+\ro_{44}+2\ro_{66}+2\ro_{88}-2\,\re \ro_{24}-4\, \re \ro_{68})\\
&v=\frac{1}{4}\,(-\ro_{23}-\ro_{27}+\ro_{43}+\ro_{47}+2\ro_{69}-2\ro_{89})
\end{split}\label{roasdetails}
\end{equation}
and
$$
t=1-2x-2y
$$
Depending on the initial state, the asymptotic state (\ref{roas}) can be separable or entangled.
In the next section we will study distillability of  $\ras$ for some initial states.
\section{Generation of stationary distillable entanglement}
As was shown in \cite{DJ1}, the negativity of the asymptotic states (\ref{roas}) can be obtained
analytically in the case of diagonal (i.e. separable) initial states. For such states, only the parameters $x,\, y$ and $t$
are non-zero and the asymptotic negativity reads
\begin{equation}
N(\ras)=\frac{1}{2}\, \left[\, \sqrt{4(x^{2}+y^{2})+t^{2}}-t\,\right]\label{negas}
\end{equation}
Note that every nontrivial asymptotic states from that class is entangled. Now we show that
this entanglement is free i.e. all such asymptotic states  are distillable. To do this, we show that
$\ras$ corresponding to diagonal initial states, violate reduction criterion (\ref{red}). Indeed, since for such states
\begin{equation}
\ptr{B}\ras\otimes\I-\ras=
\begin{pmatrix}x&0&0&0&0&0&0&0&0\\
0&x&0&0&0&0&0&0&0\\
0&0&0&0&0&0&x&0&0\\
0&0&0&y&0&0&0&0&0\\
0&0&0&0&y&0&0&0&0\\
0&0&0&0&0&0&0&y&0\\
0&0&x&0&0&0&a&0&0\\
0&0&0&0&0&y&0&b&0\\
0&0&0&0&0&0&0&0&c
\end{pmatrix}\label{redA}
\end{equation}
where
$$
a=1-2x-y,\quad b=1-x-2y,\quad c=x+y
$$
and the matrix on the right hand side of (\ref{redA}) has two negative
leading principal minors (other minors are positive), so
$$
\ptr{B}\,\ras\otimes \I-\ras\not\geq 0
$$
Similarly
$$
\I\otimes \ptr{A}\,\ras-\ras \not\geq 0
$$
The interesting examples of nontrivial asymptotic states are given
by  the separable initial states where the one atom is in the
excited state and the other is in the ground state or two atoms are
in different excited states. In all such cases, the created
entanglement is free and can be distilled.
\par
Now we consider the
possibility of creating free stationary entanglement from the bound
initial entanglement. As the initial states we take the family
\cite{HHH4}
\begin{equation}
\ro_{\al}=\frac{2}{7}\;\ket{\Psi_{0}}\bra{\Psi_{0}}+\frac{\al}{7}\;P_{+}+\frac{5-\al}{7}\;P_{-},\quad 3<\al\leq 4
\label{inibound}
\end{equation}
where
$$
\ket{\Psi_{0}}=\frac{1}{\sqrt{3}}\,\sum\limits_{j=1}^{3}\ket{j_{A}}\otimes\ket{j_{B}},
$$
$$
P_{+}=\frac{1}{3}\,\left(P_{\ket{1_{A}}\otimes\ket{2_{B}}}+P_{\ket{2_{A}}\otimes\ket{3_{B}}}+
P_{\ket{3_{A}}\otimes\ket{1_{B}}}\right)
$$
and
$$
P_{-}=\frac{1}{3}\,\left(P_{\ket{2_{A}}\otimes\ket{1_{B}}}+P_{\ket{3_{A}}\otimes\ket{2_{B}}}+
P_{\ket{1_{A}}\otimes\ket{3_{B}}}\right).
$$
The states (\ref{inibound}) are constructed as follows: we prepare the maximally
entangled state $\ket{\Psi_{0}}$ and add some specific noise resulting in
the mixing of $\ket{\Psi_{0}}$ with separable states $P_{+}$ and $P_{-}$. For a special choice
of mixing parameter such prepared states  have positive  partial transposition but
are entangled,  as can be shown by computing the realignment
negativity. For $\ro_{\al}$ it is given by
\begin{equation}
N_{\mr{R}}(\ro_{\al})=\frac{1}{21}\,\left(\,\sqrt{3\al^{2}-15\al+19}-1\,\right)\label{NR}
\end{equation}
and is obviously positive for $3<\al\leq 4$. So the states (\ref{inibound}) are bound entangled
and the entanglement initially present in $\ket{\Psi_{0}}$ cannot be extracted from them,
for any number of copies of the states. It is worth to notice that recently the bound entanglement
was created experimentally in the system of three qubits \cite{KBPS}.
\par
Although the states (\ref{inibound}) are not diagonal, one can check
that the corresponding asymptotic states have the same form as in
the diagonal case. In fact, for all initial states $\ro_{\al}$ there
is only one asymptotic state $\ras$ given by
$$
x=y=\frac{5}{56}\quad\text{and}\quad t=\frac{9}{14}
$$
By the above discussion, this  state is entangled and moreover its
entanglement is distillable. Thus we have shown that the physical
process of spontaneous emission in the radiatively coupled three -
level atoms can transform initial bound entanglement into free
distillable entanglement of the asymptotic state.
\section{Delayed creation of distillable entanglement}
In this section we study in details the evolution of entanglement of
the bound entangled initial states (\ref{inibound}) for $3<\al\leq
4$, beyond the strong correlation regime. In that case, the
asymptotic state is trivial, but some transient entanglement between
the atoms can be produced. For simplicity, we consider such atomic
configuration for which the cross coupling coefficients are equal to
zero. One can check that  the initial states (\ref{inibound}) will
evolve into the states of the form
\begin{equation}
\ro_{\al}(t)=\begin{pmatrix}\ro_{11}&0&0&0&\ro_{15}&0&0&0&\ro_{19}\\
0&\ro_{22}&0&0&0&0&0&0&0\\
0&0&\ro_{33}&0&0&0&\ro_{37}&0&0\\
0&0&0&\ro_{44}&0&0&0&0&0\\
\ro_{51}&0&0&0&\ro_{55}&0&0&0&\ro_{59}\\
0&0&0&0&0&\ro_{66}&0&\ro_{68}&0\\
0&0&\ro_{73}&0&0&0&\ro_{77}&0&0\\
0&0&0&0&0&\ro_{86}&0&\ro_{88}&0\\
\ro_{91}&0&0&0&\ro_{95}&0&0&0&\ro_{99}
\end{pmatrix}\label{roat}
\end{equation}
where all non-zero matrix elements are time dependent.
\par
Numerical analysis indicates that during the time evolution the
realignment negativity (\ref{rneg}) of the initial state very
rapidly goes to zero, so the system almost immediately disentangle.
To consider possible creation of free entanglement, let us first
check if  PPT condition can be violated during such evolution. After
taking the partial transposition, the state (\ref{roat}) becomes
\begin{equation}
\ro_{\al}(t)^{\mr{PT}}=\begin{pmatrix}
\ro_{11}&0&0&0&0&0&0&0&\ro_{37}\\
0&\ro_{22}&0&\ro_{15}&0&0&0&0&0\\
0&0&\ro_{33}&0&0&0&\ro_{19}&0&0\\
0&\ro_{51}&0&\ro_{44}&0&0&0&0&0\\
0&0&0&0&\ro_{55}&0&0&0&\ro_{68}\\
0&0&0&0&0&\ro_{66}&0&\ro_{59}&0\\
0&0&\ro_{91}&0&0&0&\ro_{77}&0&0\\
0&0&0&0&0&\ro_{95}&0&\ro_{88}&0\\
\ro_{37}&0&0&0&\ro_{86}&0&0&0&\ro_{99}\label{roatpt}
\end{pmatrix}
\end{equation}
One can check that determinant $d$ of the matrix (\ref{roatpt}) equals
\begin{equation}
\begin{split}
d&=(\ro_{22}\ro_{44}-|\ro_{15}|^{2})(\ro_{33}\ro_{77}-|\ro_{19}|^{2})
(\ro_{66}\ro_{88}-|\ro_{59}|^{2})\\
&\times(\ro_{11}\ro_{55}\ro_{99}-\ro_{55}|\ro_{37}|^{2}-\ro_{11}|\ro_{86}|^{2})
\end{split}\label{detroatpt}
\end{equation}
\begin{figure}[t]
\centering {\includegraphics[height=52mm]{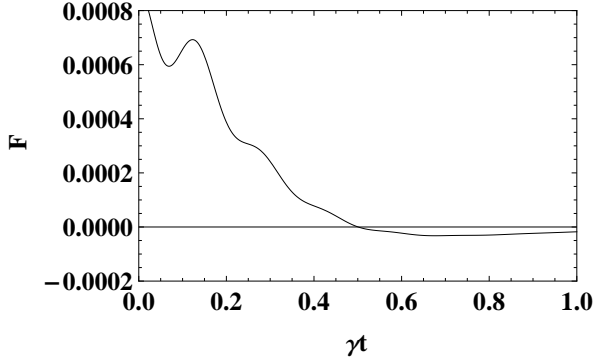}}\caption{The time evolution of the  factor $F=\ro_{11}\ro_{55}\ro_{99}-\ro_{55}|\ro_{37}|^{2}-\ro_{11}|\ro_{86}|^{2}$
 in
(\ref{detroatpt}) for the initial state (\ref{inibound}) with $\al=3.6$ and  the interatomic distance $R=0.2\,\lambda$.}
\end{figure}
We can show  numerically that (\ref{detroatpt}) changes the sign,
since the last factor is positive for all $t< t_{N}$ and becomes
negative if $t> t_{N}$, for some $t_{N}>0$, and the remaining
factors are positive. Moreover, all other leading principal minors
of the matrix (\ref{roatpt}) are always positive. So the evolution
of the bound entangled state (\ref{inibound}) has the interesting
property: for all $t< t_{N}$ the states $\ro_{\al}(t)$ are PPT and
then suddenly they become NPPT states (see FIG. 1).
\par
Now we discuss distillability of the states $\ro_{\al}(t)$. Since we
cannot exclude the possibility that there are NPPT states which are
non-distillable, we try to apply the reduction criterion of
entanglement. As we know from the discussion in Sect. II, any state
violating this criterion is necessarily distillable. By direct
computations we show that the matrix
$$
\ptr{B}\;\ro_{\al}(t)\otimes\I-\ro_{\al}(t)
$$
equals to
\begin{equation}
\begin{pmatrix}r_{11}&0&0&0&-\ro_{15}&0&0&0&-\ro_{19}\\
0&r_{22}&0&0&0&0&0&0&0\\
0&0&r_{33}&0&0&0&-\ro_{37}&0&0\\
0&0&0&r_{44}&0&0&0&0&0\\
-\ro_{15}&0&0&0&r_{55}&0&0&0&-\ro_{59}\\
0&0&0&0&0&r_{66}&0&-\ro_{68}&0\\
0&0&-\ro_{73}&0&0&0&r_{77}&0&0\\
0&0&0&0&0&-\ro_{86}&0&r_{88}&0\\
-\ro_{91}&0&0&0&-\ro_{95}&0&0&0&r_{99}
\end{pmatrix}\label{redroa}
\end{equation}
where
\begin{equation}
r_{kk}=\WW{\ro_{11}+\ro_{22}+\ro_{33}-\ro_{kk}}{k=1,2,3}{\ro_{44}+\ro_{55}+\ro_{66}-\ro_{kk}}{k=4,5,6}
{\ro_{77}+\ro_{88}+\ro_{99}-\ro_{kk}}{k=7,8,9}
\end{equation}
\begin{figure}[t]
\centering {\includegraphics[height=52mm]{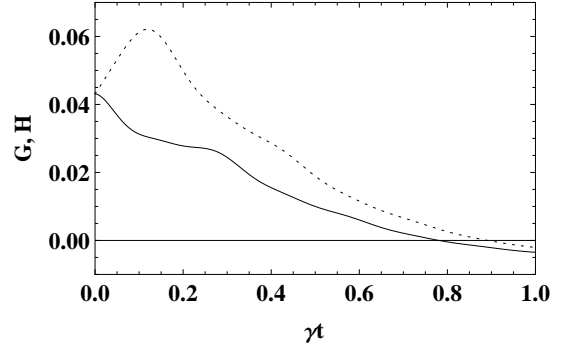}}\caption{The time evolution of the factors $G=r_{33}r_{77}-|\ro_{37}|^{2}$ (dotted line) and
$H=r_{66}r_{88}-|\ro_{68}|^{2}$ (solid line) in (\ref{mdist}) for the initial state (\ref{inibound}) with $\al=3.6$ and $R=0.2\,\lambda$.}
\end{figure}
We compute that leading principal minors of the matrix
(\ref{redroa}) which can change the sign during the evolution, and
find the following expressions
\begin{equation}
\begin{split}
m_{5}=&r_{22}r_{33}r_{44}\,(r_{11}r_{55}-|\ro_{15}|^{2})\\
m_{6}=&r_{22}r_{33}r_{44}r_{66}\,(r_{11}r_{55}-|\ro_{15}|^{2})\\
m_{7}=&r_{22}r_{44}r_{66}\,(r_{11}r_{55}-|\ro_{15}|^{2})\,(r_{33}r_{77}-|\ro_{37}|^{2})\\
m_{8}=&r_{22}r_{44}\,(r_{11}r_{55}-|\ro_{15}|^{2})\,(r_{33}r_{77}-|\ro_{37}|^{2})\,(r_{66}r_{88}-|\ro_{68}|^{2})
\label{mdist}
\end{split}
\end{equation}
where $m_{k}$ for $k=5,6,7, 8$ are  determinants of principal
$k\times k$  submatrices of the matrix (\ref{redroa}). It turns out
that the factor $r_{11}r_{55}-|\ro_{15}|^{2}$ is always positive,
but as follows from the numerical analysis,
$r_{33}r_{77}-|\ro_{37}|^{2}$ as well as
$r_{66}r_{88}-|\ro_{68}|^{2}$ change the sign during the evolution
(see FIG. 2).
Let $t_{D}$ be  the time at which the factor $
r_{66}r_{88}-|\ro_{68}|^{2}$ changes the sign. We see that
$t_{D}>0$, so only after that time the matrix (\ref{redroa})
becomes non - positive. It means that for $t>t_{D}$, the states
$\ro_{\al}(t)$ are necessarily distillable. One can check
 that $t_{D}>t_{N}$, and we see that the initial bound entangled state
(\ref{inibound})   evolves in the remarkable way: for all $t\leq
t_{N}$ it is PPT, for $t_{N}<t\leq t_{D}$ it is NPPT but a priori
can be non-distillable and only after $t_{D}$ it becomes
distillable. To show this in the explicit way,
let us introduce the measure of  the violation of  reduction criterion, defined as
\begin{equation}
N_{\mathrm{red}}(\ro)=\max\,\left(\,0,\,-\,\lambda_{\mathrm{min}}^{\mathrm{red}}\,\right)\label{redneg}
\end{equation}
where $\lambda_{\mathrm{min}}^{\mathrm{red}}$ is the minimal eigenvalues of the matrix
$$
\ro_{\mathrm{red}}=\ptr{B}\,\ro\otimes\I-\ro
$$
The quantity (\ref{redneg}) can be called the reduction negativity of the state $\ro$.
For the bound entangled initial state (\ref{inibound})  the evolution of negativity and reduction negativity is
given below (FIG. 3).
\begin{figure}[t]
\centering {\includegraphics[height=52mm]{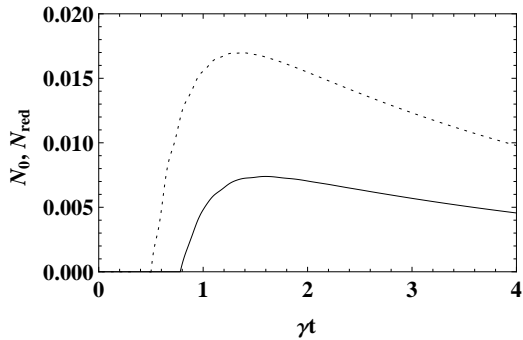}}\caption{The time evolution negativity $N_{0}$ (dotted line) and
reduction negativity $N_{\mathrm{red}}$ (solid line) for the initial state (\ref{inibound}) with $\al=3.6$ and $R=0.2\,\lambda$.}
\end{figure}
So we observe in the system the phenomenon of delayed
sudden birth of distillable entanglement. The physical reason for the appearance  of this phenomenon
can be explained as follows. During the time evolution of the system,
the process of the photon exchange produces coherence between the states
$\ket{1_{A}}\otimes \ket{3_{B}}$ and $\ket{3_{A}}\otimes\ket{1_{B}}$, so the value of $|\ro_{37}|$
starts to grow. Similarly, the same process
causes the production of coherence between states $\ket{2_{A}}\otimes\ket{3_{B}}$
and $\ket{3_{A}}\otimes\ket{2_{B}}$, so $|\ro_{68}|$ also grows. Notice that non - zero value of
$|\ro_{37}|$  or $|\ro_{68}|$ is necessary for the possibility of creation of distillable
entanglement. But this condition is not sufficient since the populations of states
of two-atomic system also evolve in time. We see from the formula (\ref{mdist}) that there is
a threshold for the reduction negativity (\ref{redneg}) at which the system becomes distillable.

The numerical value of
$t_{D}$ depends on the choice of the parameter $\al$ and the
interatomic distance $R$. For the initial state  with $\al=3.6$ and
 the distance $R=0.2\,\lambda$, we obtain $t_{D}\g\approx 0.78$ whereas $t_{N}\g\approx
0.49$.
\section{Conclusions}
We have studied the dynamics of entanglement in the system of
three-level atoms in the $V$ configuration, coupled to the common
vacuum. In the case of small (compared to the radiation wavelength)
separation between the atoms, the system has nontrivial asymptotic
states which can be entangled even if the initial states were
separable. For the large class of separable initial states the
asymptotic states are not only entangled but also distillable. The
same is true for some class of bound entangled initial states. Thus
we have shown that the dynamics of the system can transform bound
entanglement into the free distillable entanglement of stationary
states. For the atoms separated by larger distances only some
transient entanglement can exist but still the dynamical generation
of entanglement is possible. We have shown that this happens also
for the  class of bound entangled initial states. Moreover we have
demonstrated that such states evolve in a very peculiar way: they
almost immediately disentangle after the atoms begin to interact
with the vacuum, then for some finite period of time there is no
entanglement and suddenly at some time the entanglement starts to
build up. But this entanglement a priori can be  nondistillable. We
have analysed this problem using the reduction criterion of
separability and  found that the free entanglement surely appears in
the system after some additional period of time.


\begin{thebibliography}{99}
\bibitem{J} L. Jak\'{o}bczyk, J. Phys. A \textbf{35}, 6383(2002);
\textbf{36}, 1537(2003), Corrigendum.
\bibitem{FT1} Z. Ficek, R. Tana\'{s}, J. Mod. Opt. \textbf{50},
2765(2003).
\bibitem{FT2} R. Tana\'{s}, Z. Ficek, J. Opt. B, \textbf{6},
S90(2004).
\bibitem{JJ1} L. Jak\'{o}bczyk, J. Jamr\'{o}z, Phys. Lett. A
\textbf{318}, 318(2003).
\bibitem{Ag} G. S. Agarwal, \textit{Quantum Statistical Theories of
Spontaneous Emission and Their Relation to Other Approaches}
(Springer, Berlin, 1974).
\bibitem{FS} Z. Ficek and S. Swain, \textit{Quantum Interference and
Coherence: Theory and Experiments} (Springer, Berlin, 2005).
\bibitem{AP} G. S. Agarwal, A. K. Patnaik, Phys. Rev. A \textbf{63},
043805(2001).
\bibitem{EK} J. Evers, M. Kiffner, M. Macovei and Ch. H. Keitel,
Phys. Rev. A \textbf{73}, 023804(2006).
\bibitem{SE} S.I. Schmid and J. Evers, Phys. Rev. A \textbf{77},
013822(2008).
\bibitem{P} A. Peres, Phys. Rev. Lett. \textbf{77}, 1413(1996).
\bibitem{HHH1} M. Horodecki, P. Horodecki, R. Horodecki, Phys. Lett. A \textbf{223}, 1(1996).
\bibitem{HHH2}   M. Horodecki, P. Horodecki, R. Horodecki, Phys. Rev. Lett. \textbf{80}, 5239(1998).
\bibitem{YHHS} D. Yang, M. Horodecki, R. Horodecki and B. Synak-Radtke, Phys. Rev. Lett. \textbf{95}, 190501(2010).
\bibitem{DJ1} {\L}. Derkacz, L. Jak\'{o}bczyk, J. Phys. A \textbf{41}, 205304(2008).
\bibitem{DJ2} {\L}. Derkacz, L. Jak\'{o}bczyk, Phys. Lett. A \textbf{372}, 7117(2008).
\bibitem{Sun} Z. Sun, X.G. Wang, Y.B. Gao, C.P. Sun, Eur. Phys. J. D\textbf{46}, 521(2008).
\bibitem{FTd} Z. Ficek, R. Tana\'{s}, Phys. Rev. A\textbf{77}, 054301(2008).
\bibitem{B}   C.H. Bennett, G. Brassard, S. Popescu, B. Schumacher,
J.A. Smolin, W.K. Wootters, Phys. Rev. Lett. \textbf{76}, 722(1996).
\bibitem{BB} C.H. Bennett, H.J. Bernstein, S. Popescu, B.
Schumacher, Phys. Rev. A \textbf{53}, 2046(1996).
\bibitem{HHH3} M. Horodecki, P. Horodecki, R. Horodecki,Phys. Rev.
Lett. \textbf{78}, 574(1997).
\bibitem{W} J. Watrous, Phys. Rev. Lett. \textbf{93}, 010502(2004).
\bibitem{C} N.J. Cerf, C. Adami, R.M. Gingrich, Phys. Rev. A
\textbf{60}, 898(1999).
\bibitem{HH} M. Horodecki, P. Horodecki, Phys. Rev. A \textbf{59},
4206(1999).
\bibitem{VW} G. Vidal, R.F. Werner, Phys. Rev. A \textbf{65},
032314(2002).
\bibitem{H} P. Horodecki, Phys. Lett. A \textbf{232}, 333(1997).
\bibitem{HHHH} R.Horodecki, P. Horodecki, M. Horodecki, K.
Horodecki, Rev. Mod. Phys. \textbf{81}, 865(2009).
\bibitem{Ch} K. Chen, L.-A. Wu, Quantum Inf. Comput. \textbf{3},
193(2003).
\bibitem{Rud} O. Rudolph, J. Phys. A \textbf{36}, 5825(2003).
\bibitem{Rud1} O. Rudolph, Phys. Rev. A \textbf{67}, 032312(2003).
\bibitem{HHH4} P. Horodecki, M. Horodecki, R. Horodecki, Phys. Rev.
Lett. \textbf{82}, 1056(1999).
\bibitem{KBPS} H. Kampermann, D. Bruss, X. Peng and D. Suter, Phys. Rev. A \textbf{81}, 040304(R)(2010).
\end{thebibliography}
\end{document}